\documentclass[11pt,twocolumn,twoside]{osajnl}

\journal{ol} 

\setboolean{shortarticle}{true}

\title{  Breathing of dissipative light bullets of nonlinear polarization mode in Kerr resonators}

\author[1]{S. S. Gopalakrishnan}
\author[2,*]{M. Tlidi}
\author[1]{M. Taki}
\author[3,4]{K. Panajotov}

\affil[*]{Corresponding author: mtlidi@ulb.ac.be}

\affil[1]{Laboratoire de Physique des Lasers, Atomes et Mol\'{e}cules, CNRS UMR 8523, Universit\'{e} Lille 1 - 59655 Villeneuve d'Ascq Cedex, France}
\affil[2]{Facult\'{e} des Sciences, Universit\'{e} libre de Bruxelles (U.L.B), CP. 231, 1050 Brussels, Belgium}
\affil[3]{Department of Applied Physics and Photonics (IR-TONA), Vrije Universiteit Brussels, Pleinlaan 2, 1050 Brussels, Belgium}
\affil[4]{Institute of Solid State Physics, Bulgarian Academy of Sciences, Sofia, Bulgaria}

\begin{abstract}
We demonstrate the existence of breathing dissipative light bullets in a birefringent optical resonator filled with Kerr media. The propagation of light inside the cavity for each polarized component, which is coupled by cross-phase modulation, is described by the coupled Lugiato-Lefever equations. The space-time dynamics of breathing light bullets are described using Stokes parameters and frequency spectra.
\end{abstract}

\setboolean{displaycopyright}{true}

\begin{document}

\maketitle
Broad area optical resonators offer an ideal device for stdying self-organization and soliton formation in complex nonlinear systems  \cite{lugiato1994transverse,Rosanov02,kivshar2003optical,staliunas2003transverse,Ackemann2009FundamentalsAA,tlidi2016nonlinear}. 
Kerr resonators, in particular, support the recently reported dissipative light bullets (DLBs) \cite{PhysRevLett.126.153902,tlidi2021optical}. They refer to the confinement of light in two-dimensional (2D) transverse coordinates and along with a one-dimensional (1D) longitudinal direction due to diffraction and dispersion, respectively. 
They are known as light bullets in conservative systems \cite{Silberberg1,ponomarenko2006linear}. They appear to be self-organized peaks traveling at the group velocity of light. They have been theoretically predicted in dissipative systems like optical resonators filled with Kerr media \cite{tlidi19983d,Brambilla1,PhysRevLett.126.153902}, quadratic media \cite{staliunas1998three,Tlidi3,Panoiu1,Veretenov2}, saturable absorber 
\cite{vladimirov1999numerical,veretenov2016rotating}, twisted waveguide arrays \cite{Torner1}. They are solutions of the complex cubic-quintic Ginzburg-Landau equation \cite{grelu2005light,Malomed08CGL}. They have been recently reported in discrete systems such as arrays of coupled nonlinear optical cavities \cite{panajotov2021discrete}. 

\begin{figure}
 \unitlength=14.0mm
\centerline{
\includegraphics[width=3.0\unitlength,height=2.0\unitlength]{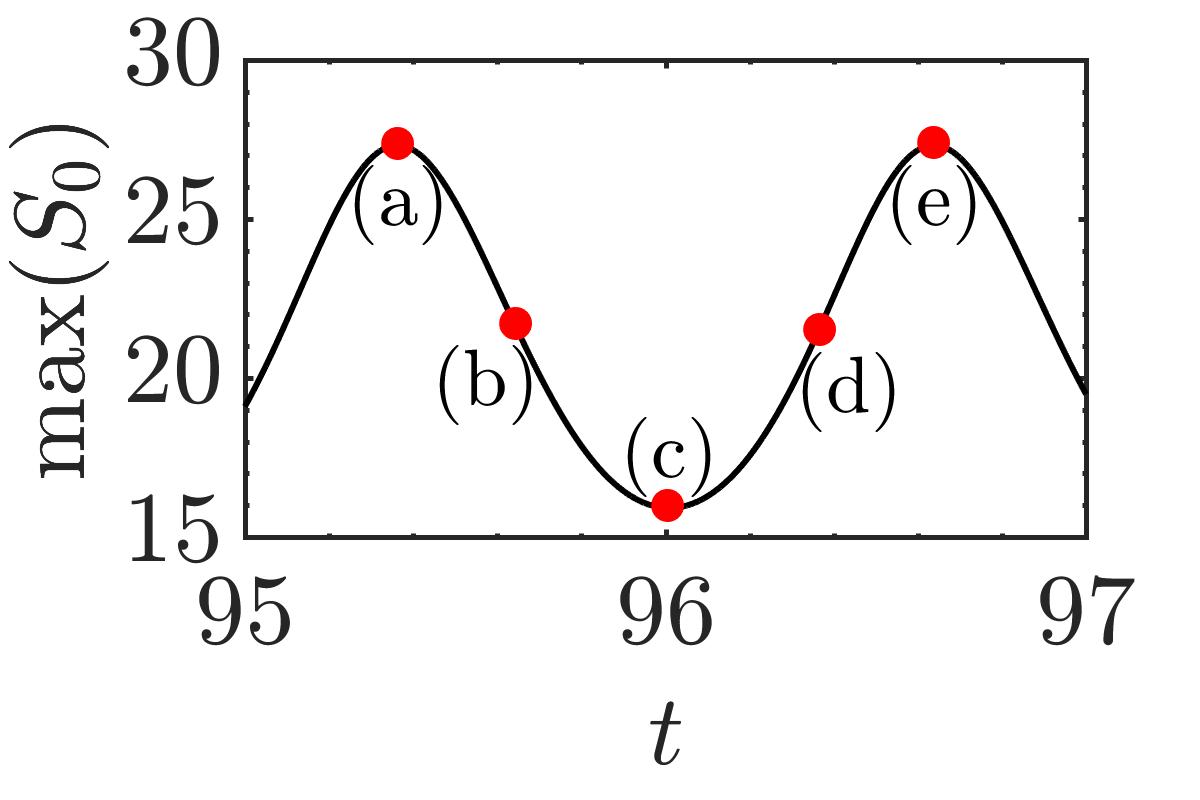}
\includegraphics[width=3.0\unitlength,height=2.0\unitlength]{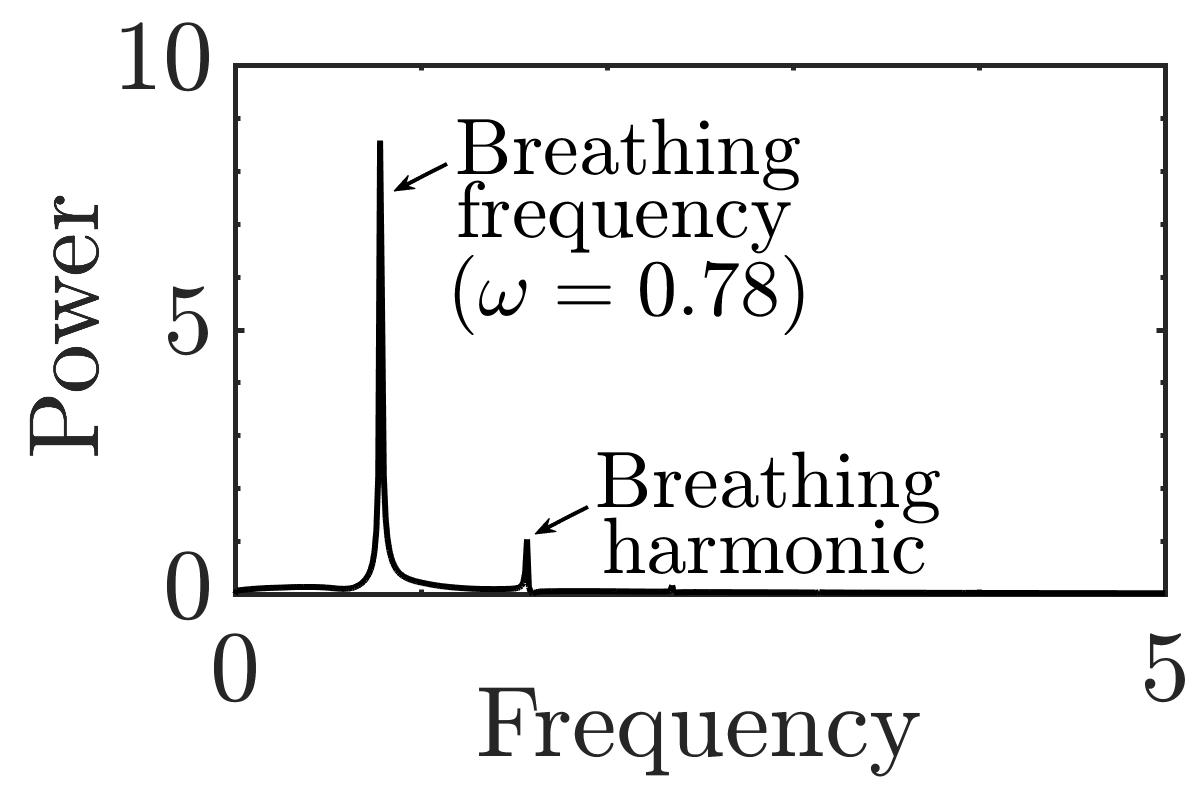}
}
 \unitlength=24.0mm
\centerline{
\includegraphics[width=3.5\unitlength,height=0.7\unitlength]{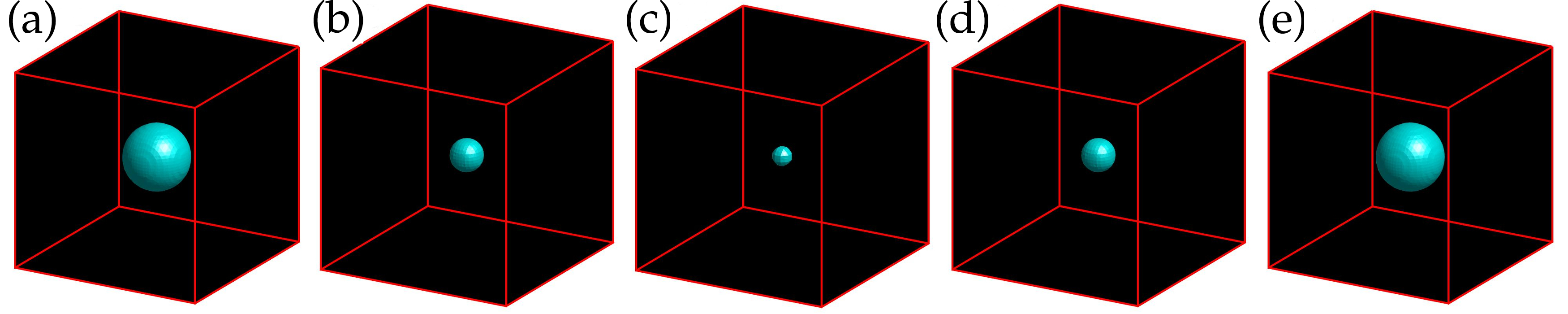}
}
 \unitlength=12.0mm
\begin{picture}(0,0)
\put(4.0,-0.2){\includegraphics[width=0.6\unitlength,height=0.7\unitlength]{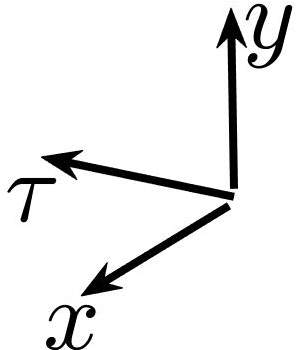}}
\end{picture}
\begin{picture}(0,0)
\put(-0.0,4.1){{(i)}}
\put(3.35,4.1){{(ii)}}
\end{picture}
 \caption{Breathing dissipative light bullet. (i) The maximum total intensity  $S_0$ within the cavity over a breathing period. (ii) The corresponding Fourier transform {of the total intensity} indicating the breathing  frequency. The higher harmonics is also present in the frequency spectra. The 3D isosurfaces represented using $1\%$ of the maximum intensity at time instants (a--e) are shown below. Parameter settings: $E_I= 2.7$, $\theta_x = 5$, $\theta_y = 4$, and $\psi = \pi/4$.} 
 \label{fig:Fig1}
 \end{figure}
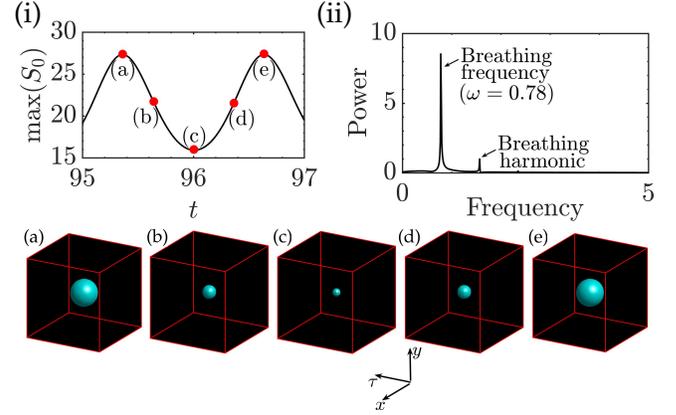

{In this Letter, we demonstrate evidence of breathing light bullets in driven optical Kerr resonators. They are caused by the polarization features of nonlinear polarisation mode interaction in Kerr optical resonators that are birefringent.}  An example of breathing DLB is shown in Fig. \ref{fig:Fig1}. We characterize the DLBs breathing, by plotting their  { Stokes parameters} and by analyzing the frequency spectra. 

{The Lugiato-Lefever equation (LLE, \cite{Lefever1}) has become a popular model for understanding the dynamics of  electric field confined in nonlinear optical resonators (see an overview on the LLE \cite{Chembo2017theory}). When considering birefringent Kerr resonators subject to a coherent optical injection, the space-time dynamics is described by the coupled  dimensionless LLE  \cite{haelterman1994polarization,Averlant1} }
\begin{eqnarray}
&&\frac{\partial E_{x,y}}{\partial t}=-\left[1+\dot{\imath}(\theta_{x,y}-|E_{x,y}|^2 - b |E_{y,x}|^2)\right]E_{x,y}+\nonumber \\
&&E_{Ix,Iy}+\left[\dot{\imath}\left(\nabla_{\perp}^2+ \frac{\partial^{2}}{\partial \tau^{2}}\right)
\pm \Delta\beta_1\frac{\partial}{\partial\tau}\right] E_{x,y}
 \label{eqn:LLE}
\end{eqnarray}
Here $E_{x,y}$ are the slowly varying electric field envelopes polarized along the slow and the fast axis, respectively.  $\theta_{x,y}$ are the frequency detunings between the  driving external fields $E_{Ix}$, $E_{Iy}$ and the cavity resonances in the corresponding polarization directions. The nonlinear cross-phase modulation coefficient is $b=2/3$ corresponding to an isotropic Kerr medium. The time $t$ corresponds to the slow-time evolution of $E_{x,y}$ over successive round-trips, whereas $\tau$ accounts for the fast time in a reference frame travelling at the group velocity of light in the Kerr medium. The injected beam amplitudes are $(E_{Ix},E_{Iy})=(\cos{\psi},\sin{\psi})E_I$, with $\psi$ the angle of the linear polarization direction of $E_I$ with respect to the orientation of the slow axis. The diffraction is described by the Laplace operator $\nabla_{\perp}^2= \partial_{xx} + \partial_{yy}$ acting on the transverse plane $(x, y)$. The chromatic dispersion  is described by  $\partial^{2}/\partial \tau^{2}$. The group-velocity mismatch associated with the first order dispersion is denoted by $\Delta\beta_1$.

{The relationships between physical variables and parameters and dimensionless variables and parameters are:  $E_{x,y}\rightarrow(\kappa/\gamma l)^{1/2}E_{x,y}$, $E_{Ix,Iy}\rightarrow \kappa(\kappa/\gamma \theta l)^{1/2} E_{Ix,Iy}$, with $\kappa$ the total losses, $\gamma$ the nonlinear coefficient $\gamma =  n_2 \omega_0/(cA_{\textrm{eff}})$ with $n_2$ being the nonlinear refractive index of the Kerr material considered, 
where $\omega_0$ is the injected field frequency,  $c$ is the speed of light, and  $A_{\textrm{eff}}$ is the effective surface illuminated in the transverse plane. The detuning parameters are $\theta_{x,y}=\phi_{x,y}/ \kappa$  where $\phi_{x,y}$  is the linear phase shift accumulated by the intracavity fields over the cavity length $l$.   The time and space coordinates in the physicall units are $(x,y)\rightarrow [l/(2 q \kappa)]^{1/2}(x,y)$ and $(t,\tau) \rightarrow [(t_r/\kappa)^{1/2}t, (\beta_2l/(2\kappa))^{1/2}\tau]$  where $t_r= l n/c$ is the round trip time, $\beta_2$ denotes the second-order chromatic dispersion coefficient, and $q$ is the wavenumber. We propose to use chalcogenide glass as optical materials that filled the optical resonator:  $\gamma\approx 0.144 \,\, \textrm{W}^{-1}\textrm{km}^{-1}$ for  $n_2\approx 2.3\times 10^{-17}\,\textrm{m}^2/\textrm{W}$, $A_{\textrm{eff}}=25\times10^4$  $\mu \textrm{m}^2$, $l=1000$ $\mu\textrm{m}$, $\kappa  \approx 40\,\textrm{dB/km}$, the reflectivity of the mirrors $\theta=0.95$.   With these realistic physical parameters taken from  \cite{snopatin2009high,zakery2007optical}, the intensity of the injected field is of the order of $10$ $\textrm{MW}/\textrm{cm}^2$, the spatial width in $(x,y)$ plane will be $\approx 270 \,\,\mu\textrm{m}$ and the temporal width of the DLB is $\approx 0.08\,\textrm{ps}$ for a value of $\beta_2 \approx 20\, \textrm{ps}^2/\textrm{km}$ and  the  group-velocity mismatch $\Delta\beta_1 = 0.1$ which corresponds to a physical value $\Delta\beta^\prime_1 \approx 244~ \textrm{ps}/\textrm{km}$. }

The  Kerr resonators support stationary localized structures in 1D and 2D settings  \cite{scroggie1994pattern,firth2002dynamical}, as well as breathing of localized structures  \cite{VladimirovPRL2012}, in the scalar case where polarisation degree of freedom is ignored.  However,  in one dimesnional setting, the stability of localized structures are strongly dependent on the polarization of the incident light as 
they acquire a distinct ellipticity \cite{Averlant1,saha2020polarization,kostet2021vectorial,kostet2021coexistence}. These dissipative localized structures can undergo breathing, which is a periodic variation in their duration and intensity \cite{Matsko12,Bao2016,yu2017breather,Kippenberg1}.

In order to investigate the dynamics of a single  DLB, we use as an initial condition a Gaussian function { with amplitude one and half width of $0.3$} added to the value of the continuous wave (CW) centred within the optical cavity.  {The  CW solutions of Eq. (1)  satisfy $I_{Ix,Iy}= [1+(\theta_{x,y}-I_{x,y} - 2I_{y,x}/3)^2]I_{x,y}$, with $I_{Ix,Iy}=E_{Ix,Iy}^2$ and $I_{x,y}=|E_{x,y}|^2$ are the strengths of the injected and intracavity fields, respectively. The frequency detuning parameters  are set to $\theta_x = 5$ and $\theta_y = 4$. The CW solutions show tristability in this example.
Except for the lowest CW solution, the linear stability analysis of the CW solutions with respect to a finite frequency perturbation suggests that all of the CW solutions are modulationally unstable. } The spatial discretization is done using a 
Fourier spectral method with periodic boundary conditions \cite{Kassam2}, and the time-stepping is carried out with a fourth order exponential time differencing Runge-Kutta scheme \cite{Matthews1}. All the numerical simulations in the present study are carried out using a time-step of $0.01$ on a periodic domain of size $80$ units along $(x,y,\tau)$ directions resolved using $128$ grid points.  

Numerical simulations of Eq. \ref{eqn:LLE} with periodic boundary conditions along the $(x,y,\tau)$ directions indicate that a single dissipative light-bullet exhibits a breathing dynamics as shown in Fig. \ref{fig:Fig1}. The maximum of the intracavity field intensity
$S_0 = \vert E_x\vert^2 + \vert E_y\vert^2$ is plotted as a function of the fast time over a breathing period as shown in  Fig. \ref{fig:Fig1}(i). A sequence of 3D isosurfaces over a breathing cycle illustrating the periodic variation of the DLB diameter is shown in Fig. \ref{fig:Fig1}(a-e). 
This behavior  evidences a typical breathing signature of DLB with a periodic variation. This is further revealed by its spectrum; the Fourier transform {of the total intensity} is shown in panel (ii) indicating the fundamental breathing frequency. The higher harmonic is also manifestly evident revealing the nonlinear interaction within the system. 
Similar breathing dynamics in 1D settings have been recently experimentally demonstrated in microresonators \cite{Kippenberg1}. 
{ By increasing the integration time, douling the domain of size, and reducing the space and time steps, we have quantitatively verified that breathing dynamics is a robust phenomenon.} 
\begin{figure}[t]
\centering
 \unitlength=20.0mm
\centerline{
\includegraphics[width=1.967\unitlength,height=1.112\unitlength]{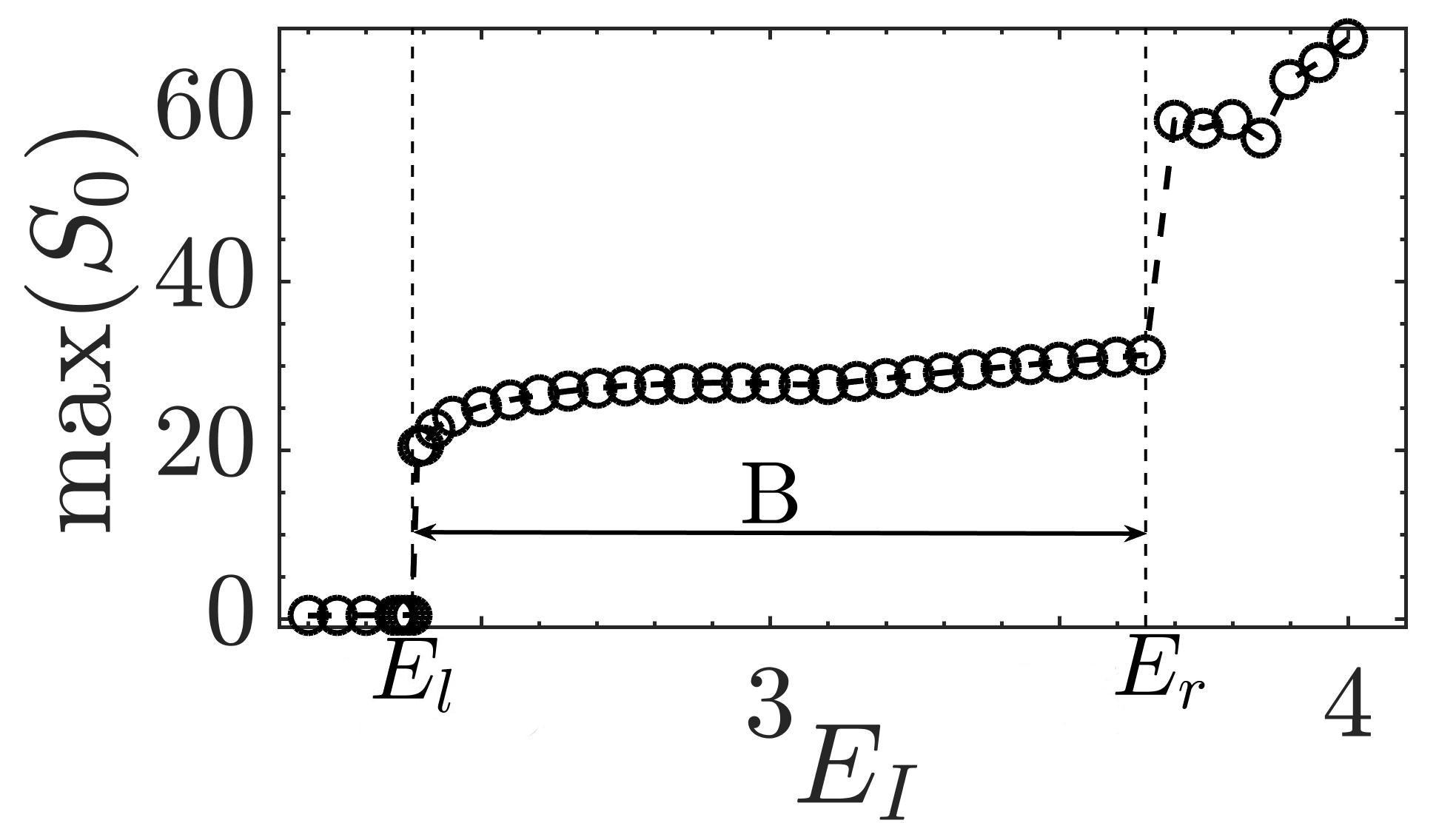}
\includegraphics[width=1.967\unitlength,height=1.112\unitlength]{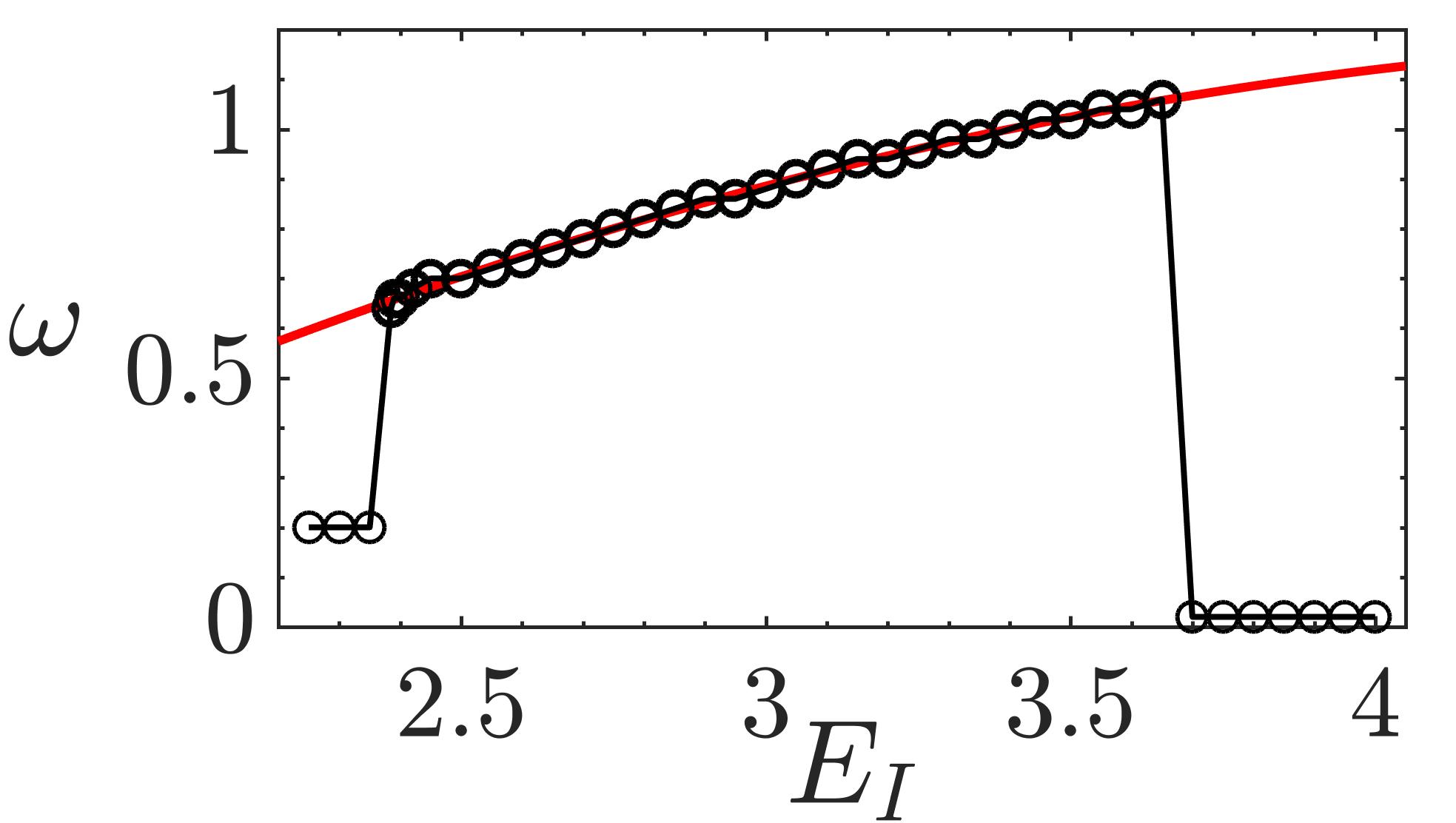}
}
\begin{picture}(0,0)
\put(-2.1,1.3){{(a)}}
\put(0.0,1.3){{(b)}}
\end{picture}
\vspace{-0.75cm}
 \caption{Effect of the optical injection. (a) Variation of the maximum total intensity $S_0$ with the injected field strength $E_I$. (b) The corresponding breathing frequency $\omega$ exhibiting a parabolic dependence with $E_I$  {shown by red line}. The breathing regime, $E_l < E_I < E_r$, is denoted by $\textrm{B}$ with $E_l = 2.38$ and $E_r =3.65$. Parameters are $\theta_x = 5$, $\theta_y = 4$, and $\psi = \pi/4$.}

 \label{fig:Fig2}
 \end{figure}
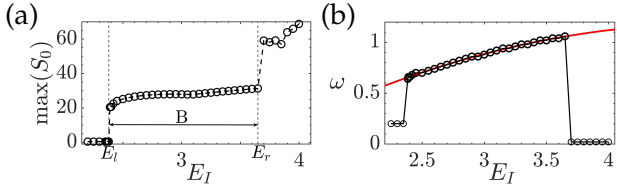 

To quantify the breathing phenomenon we vary the injected field strength while fixing the remaining system parameters. The results of numerical simulations of  Eq. \ref{eqn:LLE} are summarized in Fig. \ref{fig:Fig2}. In Fig. \ref{fig:Fig2}(a) the maximum total intensity $S_0$ is plotted  as a function of $E_I$. The corresponding frequency of the breathing DLB is shown in panel (b). For $E_l < E_I < E_r$ robust time-periodic DLBs are observed in the breathing regime, denoted by $\textrm{B}$. The fundamental breathing frequency of the DLBs fits a parabolic dependence with $E_I$  { shown by red line in \ref{fig:Fig2}(b).  Such dependence has been} observed in the one-dimensional setting \cite{Kippenberg1}. 
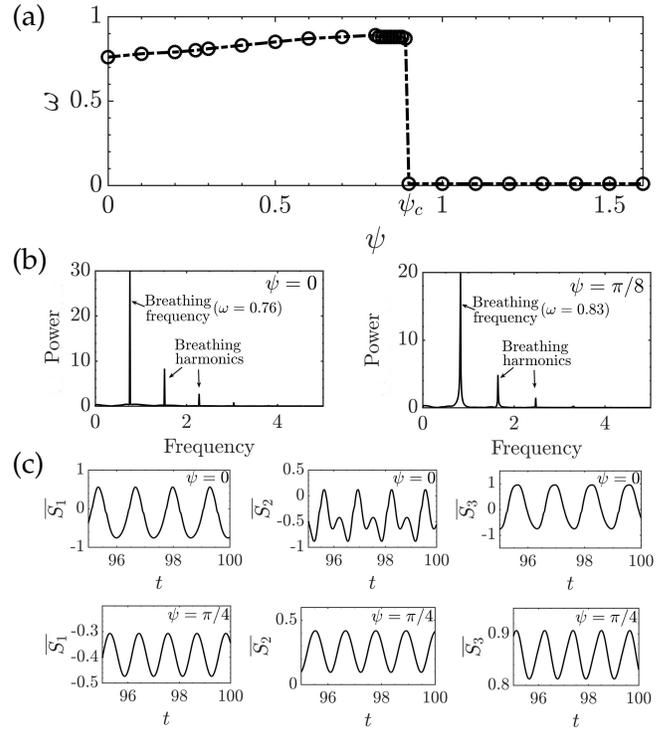
\begin{figure}[t]
\centering
 \unitlength=34.0mm
\centerline{
\includegraphics[width=2.4\unitlength,height=2.85\unitlength]{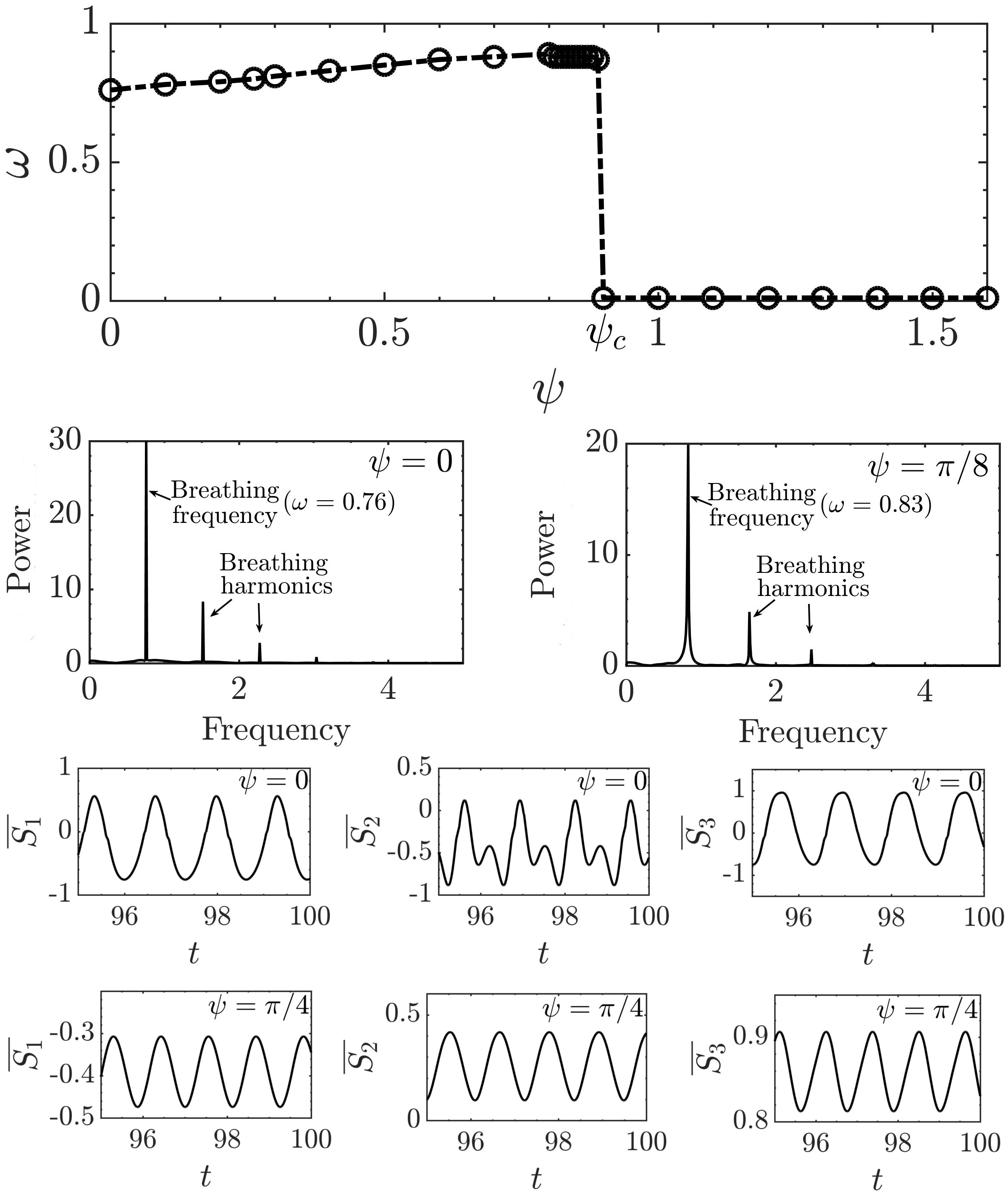}
}
\begin{picture}(0,0)
\put(-1.3,2.9){{(a)}}
\put(-1.3,1.95){{(b)}}
\put(-1.3,1.15){{(c)}}
\end{picture}
\vspace{-0.75cm}
 \caption{Effect of the angle $\psi$ on the breathing dynamics. (a) The breathing frequency $\omega$ as a function of $\psi$ revealing the critical value of $\psi$ beyond which breathing DLBs cannot be observed ($\psi_c \approx 0.9$). (b) Representative frequency spectra for $\psi =0$ and $\psi = \pi/8$. (c) Mean value of the normalised Stokes parameters within the optical cavity for $\psi=0$ and $\psi=\pi/4$ where ${\bar{S_1}}=(|E_x|^2-|E_y|^2)/S_0$, ${\bar{S_2}}=2\textrm{Re}(E_xE_y^*)/S_0$, and ${\bar{S_3}}=-2\textrm{Im}(E_xE_y^*)/S_0$. Here ($*$) stands for the complex conjugate. Parameter settings: $E_I = 3$, $\theta_x = 5$, and $\theta_y = 4$.}
\label{fig:Fig4}
\end{figure} 
 To further characterize the stability domain of DLBs, we fix the injection strength and vary the linear polarization direction with respect to the orientation of the slow axis. The results are summarized in  Fig.  \ref{fig:Fig4}. From this figure, we see that beyond a critical value of $\psi$, robust breathing DLBs ceases to exist in the system.  To illustrate the time-dependent behavior of the DLB polarization state we use the normalized Stokes parameters, which signify the amount of flux radiated in space along different orientations and with linear and circular polarization. Fig. \ref{fig:Fig4}(c) shows the periodic variation of the mean value of the normalized Stokes parameters within the cavity for two different values of $\psi=0$ and  $\psi=\pi/4$. For $\psi=0$ (Fig. \ref{fig:Fig4}(c) upper panel), ${\bar{S_1}}$ varies with time from almost $-1$ to about $0.8$, i.e. from almost $y$ to almost $x$ linear polarization. At the same ${\bar{S_2}}$ oscillates from about $-1$ to only about $0$, i.e. from linear polarization along $-\pi/4$ to no component of linear polarization along the $\pm\pi/4$ directions. Finally, ${\bar{S_3}}$ also displays large time-oscillations from about $-1$ to about $1$, i.e. left and right circular polarizations are present most of the time. This behaviour changes drastically however for $\psi=\pi/4$ (Fig. \ref{fig:Fig4}(c) bottom panel). In this case, the oscillations of the normalized Stokes parameters are much decreased with ${\bar{S_1}}$ (${\bar{S_2}}$) always negative (positive) and ${\bar{S_3}}$ large and positive, i.e. light remains almost right circularly polarized over a breathing cycle.

 \begin{figure}
\centering
 \unitlength=35.0mm
\centerline{
\includegraphics[width=1.2\unitlength,height=0.8\unitlength]{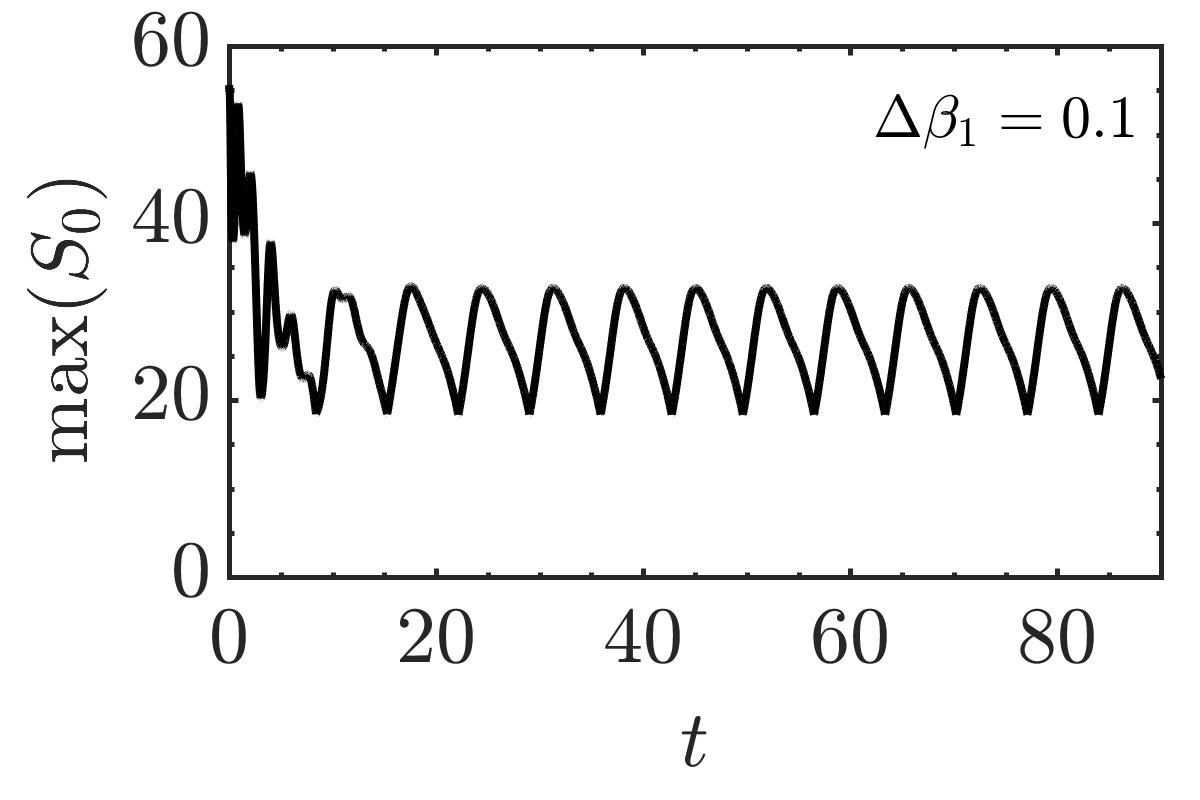}
\includegraphics[width=1.2\unitlength,height=0.8\unitlength]{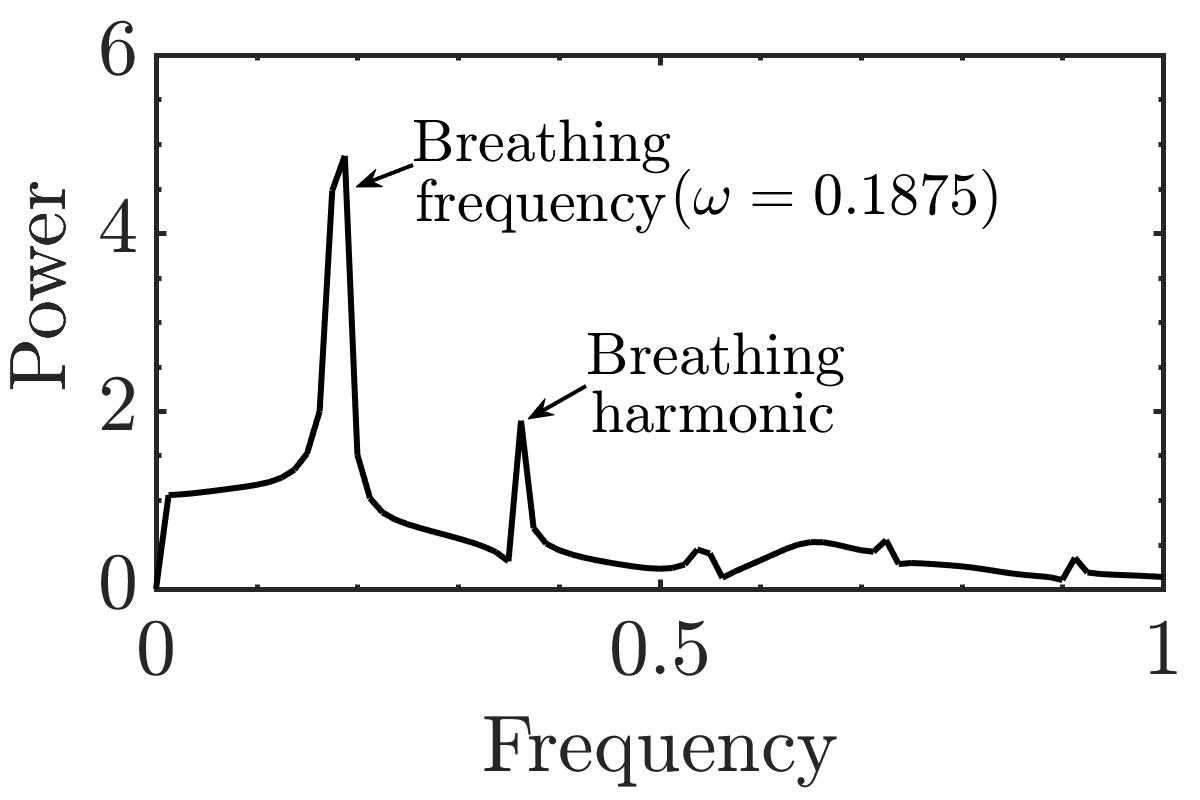}
}
 \unitlength=50.0mm
\centerline{
\includegraphics[width=0.8\unitlength,height=0.8\unitlength]{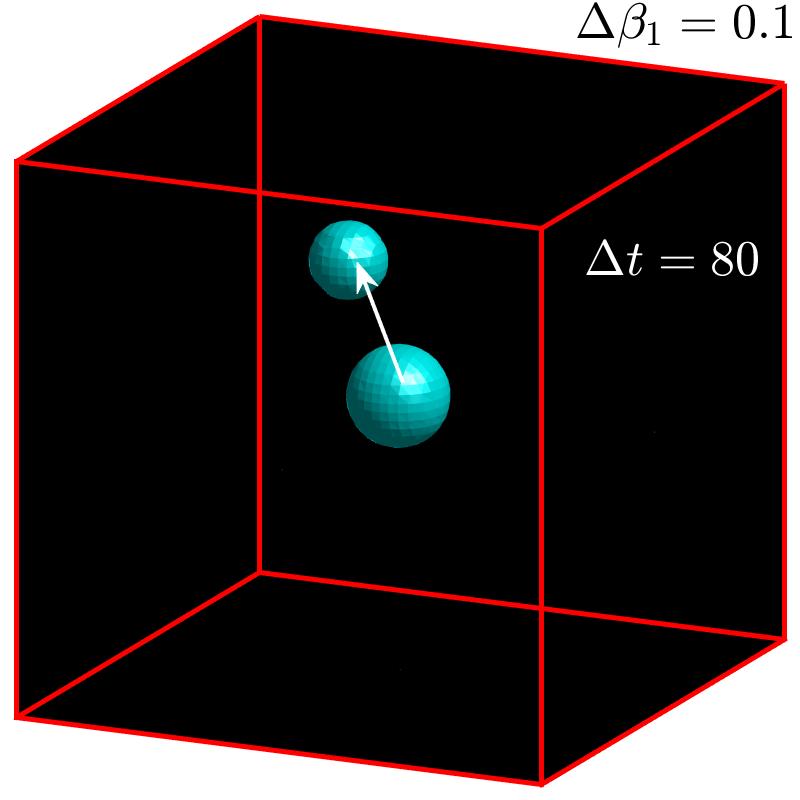}
\includegraphics[width=0.8\unitlength,height=0.8\unitlength]{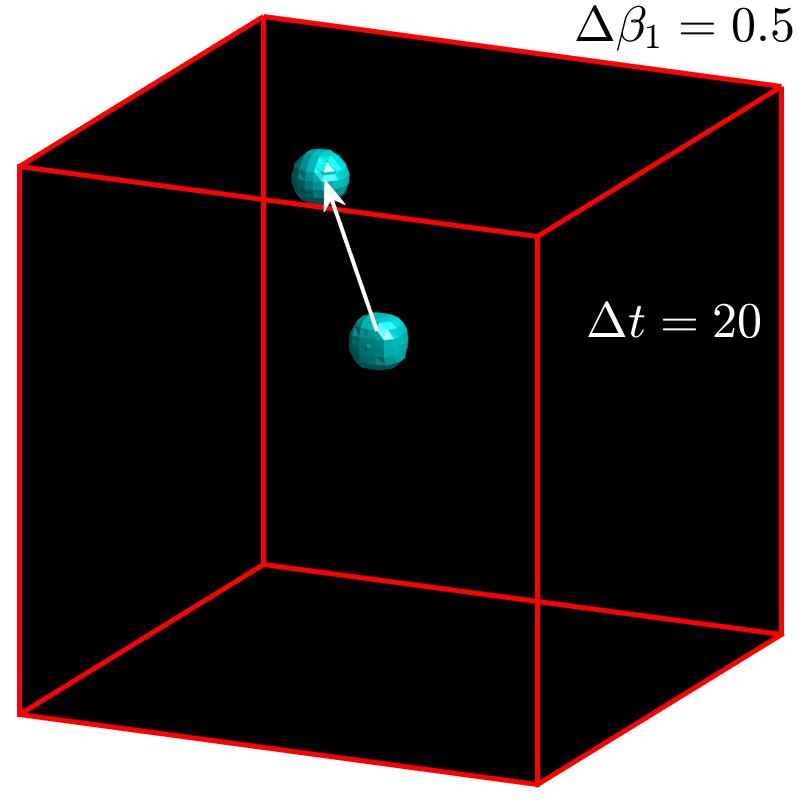}
}
\begin{picture}(0,0)
\put(-0.9,1.42){{(a)}}
\put(-0.02,1.42){{(b)}}
\put(-0.9,0.85){{(c)}}
\put(0.0,0.85){{(d)}}
\end{picture}
\begin{picture}(0,0)
\put(-0.18,-0.05){\includegraphics[width=0.15\unitlength,height=0.175\unitlength]{Figures/Fig1d.jpg}}
\end{picture}
 \caption{Effect of group-velocity mismatch ($\Delta\beta_1 \neq 0$). The breathing phenomenon is still present albeit the DLB starts to drift, as a function of $\Delta\beta_1$. Parameter settings: $E_I = 3.0$, $\theta_x = 6.5$, $\theta_y = 4.5$, and $\psi = \pi/4$ (a,b,c) $\Delta\beta_1 = 0.1$, (d) $\Delta\beta_1 = 0.5$.}
 \label{fig:Fig5}
 \end{figure} 

{ The breathing DLB shown bive have been generated in the limit of the zero group velocity mismatch.
The impact of the group velocity mismatch can be significant.
At two distinct values of $\Delta \beta_1=(\beta_{1x}-\beta_{1y})/2$, this impact is demonstrated in Fig. \ref{fig:Fig5}.
The DLBs keep their breathing nature, but when $\Delta \beta_1 \neq 0$, the symmetry along the $\tau$ direction is broken, the DLB drifts in the reference frame at the velocity that is the mean group velocity of the two components. 
The velocity with which the DLB drifts rises as $\Delta \beta_1$ increases.  The change of position of the DLB in $(x,y,\tau)$ is showm with arrows in Fig.  \ref{fig:Fig5}(c,d).}

\noindent\textbf{Disclosures.} \small{The authors declare no conflicts of interest.}\\

\noindent\textbf{Acknowledgements.}
\small{
We acknowledge the support from the French National Research Agency as well as the French Ministry of Higher Education and Research, Hauts de France council and European Regional Development Fund (ERDF) through the Contrat de Projets Etat-Region (CPER Photonics for Society P4S). M. T. acknowledges financial support from the Fonds de la Recherche Scientific FNRS under Grant CDR No. 35333527 ``Semiconductor optical comb generator''. 
This work was supported in part by the Fonds Wetenschappelijk Onderzoek-Vlaanderen FWO (G0E5819N).}

\bibliography{BDLB}
\bibliographyfullrefs{BDLB}

\bibliographyfullrefs{Biblio_LL_LB_dis_red_v2}

\end{document}